# Inflection Point: A New Perspective on Photonic Nanojets

*Guoqiang Gu, Pengcheng Zhang, Sihui Chen, Yi Zhang and Hui Yang*[*]

Dr. G. Gu, Dr. P. Zhang, Mr. S. Chen, Dr. Y. Zhang, Dr. H. Yang
Laboratory of Biomedical Microsystems and Nano Devices
Bionic Sensing and Intelligence Center
Institute of Biomedical and Health Engineering
Shenzhen Institutes of Advanced Technology
Chinese Academy of Sciences
Shenzhen 518055, China
E-mail: hui.yang@siat.ac.cn

Dr. H. Yang
CAS Key Laboratory of Health Informatics
Shenzhen Institutes of Advanced Technology
Chinese Academy of Sciences
Shenzhen 518055, China

**ABSTRACT**

When light propagates through the edge or middle part of microparticle's incoming interface, there is a basic rule that light converges and diverges rapidly or slowly at the output port. These two parts are referred to as region of rapid change (RRC) and region of slow change (RSC), respectively. Finding the boundary point between RRC and RSC is the key to reveal and expound this rule scientifically. Based on the correlation between light convergence-divergence and the slope of emergent light, combined with the relationship between natural logarithm and growth in physical reality and the second derivative of a function in practical significance, we determine the boundary point between RRC and RSC, namely the inflection point. From such perspective, photonic nanojet (PNJ) and near-field focusing by light irradiation on RSC and RRC, as well as the position of the inflection point under different refractive index contrast and the field distribution of light-focusing, are studied with finite-element-method-based numerical simulation and ray-optics-based theoretical analysis. By illuminating light of different field intensity ratios to the regions divided by the inflection point, we demonstrate the generation of photonic hook (PH) and the modulation of PNJ/PH in a new manner.

## Introduction

At the beginning of the new millennium, a phenomenon of jet-like-structured field enhancement effects originating from light-particle interaction was first known to the researchers in laser physics.[1-3] In 2004, this phenomenon was thoroughly studied using numerical analysis on dielectric microcylinders illuminated by a plane wave and formally referred to as the "photonic nanojet (PNJ)"[4]. PNJ is a non-resonant light beam, which has a subwavelength transverse beam width, high optical intensity and wavelength-scale propagation distance with weak diffraction, that emerging form the shadow-side surface of the irradiated cylindrical or spherical microparticles with dimensions comparable or slightly larger than the radiation wavelengths.[5] Over the past sixteen years, PNJs have been applied in many fields such as enhancements of backscattering, fluorescence, transmission and Raman scattering,[6-9] optical forces,[10] optical data storage,[11] coupled resonator optical waveguide,[12] nanopatterning, nanolithography and nano-marking,[13-15] super-resolution imaging,[16, 17] surface plasmon polariton,[18] scanning probe,[19] all-optical switching,[20] single nanoparticle detection,[21, 22] biomolecules/cells trapping and manipulation,[23, 24], Bloch surface waves,[25] optofluidic microcavity,[26] biosensor and micromotor,[27] backaction of microparticles,[28] etc. The shape features and field distributions of the PNJs, which are usually characterized by the parameters of maximum light intensity ($I_{max}$), focal distance (*f*) or working distance ($W_d$), transverse beam width ($\omega$), and decay length, are dependent on the material attributes and geometrical morphologies of the cylindrical or spherical microparticles, the surrounding medium, and the natures (type, wavelength, polarization, etc.) of the incident light.[5, 29] There are three basic rules followed between the key parameters (*f*, $W_d$, and $\omega$) of PNJs and the dielectric properties (refractive index, dimension and shape) of microparticles.[4, 6, 30] First, the larger refractive index of the microparticle produces PNJs of closer focal distance $f$ (or working distance $W_d$) and narrower transverse beam width $\omega$. Second, on the contrary, the larger the curvature radius of the microparticles, the further $f$ (or $W_d$) and the wider $\omega$ for the generated PNJs. Third, a tight focus, with small $\omega$ and short $f$ (or $W_d$), is formed by illuminating the two edge regions of the first light refraction interface (LRI), while a loose focus, with large $\omega$ and long $f$ (or $W_d$), is formed by illuminating the middle region of the first LRI. The research on PNJ must be based on a full understanding of these three rules. The former two rules, with clear and definite physical mechanisms, have received general attention and been extensively studied by a number of theoretical and experimental works.[31, 32]

But so far, a systematic discussion on the third rule is lacking. It has been either implicitly or explicitly involved in the studies from the last few years. For example, by decorating concentric ring structures into the first LRI of spherical microparticle with unexplained location, number and depth, Wu *et al.* designed and demonstrated the modulation of PNJs via finite-difference time-domain simulation, focus ion beam milling and optical microscope imaging.[33] By introducing circular pupil masks of variable sizes to block part or all of the light propagating to the incident plane, Yan *et al.* attained a flexible and easy-to-implement way to tune the focusing of a microsphere lens in a controllable manner.[34] A similar principle of creating non-transparent cover masks on the illumination surfaces of microspheres was used to filter the propagating beams near the optical axis and form PNJs with sharp focal spots.[35, 36] The intensity enhancement and waist shortage of the produced PNJs were anomalously achieved at the same time in the model of axially illuminated circular-column particle-lenses covered by aluminum pupil-masks.[37] One can also observe that the opaque masks mounted in front of the micro-ellipsoid/microcylinder or a cuboid embedded inside a microcylinder, with locations and sizes not following a critical physics rule, were used to partially block the incident light and control the convergence of the focusing field in the area of curved PNJ, i.e. the so-called photonic hook (PH).[38-40] By changing the major-to-minor axis ratio of the elliptical microcylinder, which can be substantially viewed as the curvatures of the points on one middle region and two edge regions of the first LRI with different rates of change, the field distribution and location of PNJs can be artificially modified.[41] Such abstract entity was also embodied, to some extent, in the research of twin PHs produced by twin-ellipse microcylinder.[42] Through the division of a circular-column particle into a horizontal graded-index microcylinder constituted by multilayer media of equal thickness along the direction of light propagation, the side-lobes-controlled ultra-narrow PNJs could be generated and modified.[43] These works have intentionally or unintentionally taken advantage of the special focusing characteristics resulting from the aforementioned rule three, it is just that none of them has delved into the partition of incoming interface in a cause-and-effect way. The following two works explicitly involved and made a preliminary exploration in this problem. First, Wang *et al.* analyzed the working distance and beam width of the focus under the illumination conditions of covering the whole LRI, localizing in the middle, two boundaries and partial areas of the microsphere lenses, and the related influences on contrast enhancement and imaging mechanism of a near-field assisted white light interferometry.[44] Second, Gu *et al.* further clearly divided the first LRI into one middle region and two edge regions, and on this basis a microparticle with special shape was designed to realize the formation of an ultra-narrow PNJ with refractive index of the dielectric overstepping the upper bound of 2.[30] A crucial unresolved problem that arises here is: where is the boundary point between the middle region (representing slow convergence and divergence of light) and the edge region (representing rapid convergence and divergence of light) of the first LRI? None of the studies above have addressed this fundamental problem, it is only an unstated and general division of the incident plane in ref. [44] and a simple trisection of the first LRI based on $\pi/3$ opening angle for the central part in ref. [30].

Ray optics combining with numerical calculation is an effective way for qualitatively and quantitatively analyzing the formation mechanism and distribution characteristics of PNJs.[29-31, 45] The light scattering of a microparticle with no dissipation can be characterized by its refractive index and size parameters, and approximated as a ray-tracing model upon Snell's law.[31, 46] In terms of Snell's law, the speeds at which light converges and diverges are closely related to the slopes of the emergent rays, where the slopes have a dependent relationship with the incidence angles of the incident light (or the positions of the incident light intersecting with the first LRI of the microparticle).[29] The slope changes of the rays, in the case of no microparticle, are zero for all the propagating photons, while the slope changes are a nonlinear functional curve of the incidence angles in the case of a microparticle within the model. In other words, the slopes of the emergent rays for the case of containing a microparticle can be regarded as positive and negative growth of slope values with respect to the case of no slope changes. The speed of light convergence and divergence can be intuitively understood as the length of time that the slopes of the outgoing rays change. This is exactly what natural logarithm means in mathematics: the natural logarithm reveals the time required to reach a certain level of growth.[47, 48] In differential calculus and differential geometry, if the curvature of a smooth plane curve changes from being concave to convex, or vice versa, there must be a "boundary point" between the segments of concave curve and convex curve.[49] We know that mathematically, this point is called the inflection point, which can be solved by finding the second derivative of the obtained logarithmic function curve.[50]

In this paper, we present an in-depth research on the third rule of the PNJs that are produced by plane-wave-illuminated circular microparticles, i.e. the research on precisely dividing the first LRI of the microparticle into three parts: region of slow change (RSC), region of rapid change (RRC) and region of weak contribution (RWC). The key finding of this research is that the boundary point between RSC and RRC, namely the inflection point in the text, can be figured out by thoroughly understanding the slope curve of emergent rays with the help of utilizing the mathematical meaning of natural logarithm function and the second derivative of the function. An infinitely long microcylinder in three dimensions is used as the research model of a circular microparticle. Rigorous two-dimensional (2D) theoretical and numerical studies are carried out by ray optics analyses and finite element method (FEM) simulations. When RSC and RRC are separately irradiated by a plane wave, long PNJ with wide $\omega$, long $f$ ($W_d$), and short near-field focusing (including one central main lobe and two side secondary lobes) with narrow $\omega$, short $f$ ($W_d$) are visually presented. As compared to the case of RWC adding to RRC, the induced relative changes of beam width $\omega$ are, respectively, 0.25% and 0.81% for central main lobe and side secondary lobes, while the values of $f$ ($W_d$) are the same for RRC and RRC plus RWC. The inflection point position changing with the refractive index contrast (*RIC*) of the microcylinder relative to its surrounding medium, the corresponding working distance and relative value of beam width at different incident wavelength, and whether the RWC exist in the conditions that the *RIC* is greater or less than the critical value ($\sqrt{2}$), are analyzed and discussed with comprehensive ray optics theoretical analyses and full-wave simulation experiments. By illuminating RSC and the composite region (RRC+RWC) with light of different intensities, PNJ and PH are formed in the cases of the two composite regions on both sides of RSC that is irradiated with plane electromagnetic (EM) waves of equal or unequal intensity. The key parameters of PNJ and PH can be easily modulated by simply adjusting the light intensity ratios between RSC and the composite region for the cases of synchronously or asynchronously changing the light intensity of the two edge regions. The results of this study could significantly promote the understanding, utilization and application of PNJs at a fresh perspective.

## Results and Discussion

The research model expressed in three-dimensional (3D) schematic diagram is shown in Figure 1a. A monochromatic plane wave, with a free space wavelength of $\lambda_0 = 400$ nm, unit intensity of $I_0 = 1$ and wave vector $\vec{k}$ propagating along *z*-axis is set as the incident light source. The electric and magnetic field components are on the *x*-axis and *y*-axis of Cartesian coordinate respectively. The microcylinder is assumed as a homogeneous, isotropic and loss-free circular-column structure with infinite length in the *y*-direction. FEM and finite-difference time-domain are the two most widely used numerical computation methods for solving the internal and external field distributions of plane-wave-illuminated microparticles,[45] here, the FEM-based COMSOL Multiphysics software package is adopted to perform the 2D full-wave simulations in the later numerical experiments. To fully absorb the outgoing EM waves and eliminate the undesired back reflections, a perfectly matched layer is chosen as the absorbing boundary condition on the two lateral sides and one end side of the simulation area. Refractive-index-dependent non-uniform meshes with fine element sizes of less than $\lambda_0/40$ are used for the whole computational domain.

Figure 1b is the schematic diagram of the research model showing in 2D sectional view. $O$ and $r$ are respective the origin of coordinates and the radius of the microcylinder. The refractive indices of the microcylinder and the environmental medium are $n_m$ and $n_e$, respectively. RSC and RRC are the two regions, as already mentioned in the introduction section, of light converging and diverging with either slow or rapid mode. Ray-tracing models based on Geometrical optics approximation show positive effect in qualitatively analyzing and calculating the beam variations of light propagating through small particles.[46, 51] By the way of ray-tracing, the regions of RSC and RRC are marked and roughly divided with the cyan-blue and magenta-pink transmission lines. The green solid lines represent a typical ray-tracing process following the law of refraction as it travels through the dielectric microcylinder. $\theta_{iem}$, $\theta_{rem}$ and $\theta_{ime}$, $\theta_{rme}$ are, respectively, the angles of incidence ($i$) and refraction ($r$) on both sides of the interfaces when light enters and leaves the microcylinder. Taking the $z$-axis of symmetry as the reference line, the range of $\theta_{iem}$ is (0°, 90°) above the $z$-axis and (-90°, 0°) below the $z$-axis. $L$ is the position of the intersection point between the circular curve of microcylinder and the ray of emergent light. $S_{ie}$, $S_m$ and $S_{oe}$ are the slopes of the ray trajectories before, during and after entering the microcylinder. Near the upper and lower vertices of the cylinder along the $x$-axis, we define a special region of RWC on each side. The reason why this region is called the weak contribution region is that the emergent light is emitted below the *z*-axis and propagates off the *z*-axis after the incident light from the upper edge refracted by the microcylinder. Hence, when the emergent light at point *L* is treated as a secondary wave, the light from these points contributes very weak for the focal spot with a focus on the *z*-axis.

The effects of week contribution, tight focus and loose focus to the illumination light in RWC, RRC and RSC can be easily seen from the ray-tracing diagram in Figure 1c. The rays represented by the cyan-blue solid lines in RSC focus a little further away from the microcylinder. In RRC, the region marked by magenta-pink solid lines, the convergent and divergent positions of the light rays are near the rear surface of the microcylinder. For the rays in RWC, the light-gray solid lines in the enlarged view of insertion, light travels as described in the previous definition. The light incident near the upper vertex of the microcylinder has its emitting-point below the *z*-axis and propagates downward away from the *z*-axis. On the contrary, for the incident light near the lower vertex of the microcylinder, the emitting-point is above the *z*-axis and the propagation direction is upward away from the *z*-axis. If only from the perspective of ray optics, the light coming from RWC has no contribution to the focal spot on *z*-axis. However, as the interaction between light and microcylinder in actual conditions is based on wave optics theory, the secondary waves with positions deviating from the *z*-axis and propagation directions away from the *z*-axis still have a weak contribution to the light-focusing of the focal spot on the *z*-axis.

According to Snell's law, the slope of emergent ray and the *x*-coordinate of point *L* in Figure 1b can be written as

$$S_{oe} = \tan[2(\theta_{rem} - \theta_{iem})] \tag{1}$$

$$L_x = r \cdot \sin(2\theta_{\text{rem}} - \theta_{\text{iem}}) \tag{2}$$

where

$$\theta_{\text{rem}} = \sin^{-1}\left[\frac{n_e}{n_m} \cdot \sin(\theta_{\text{iem}})\right] \tag{3}$$

Here, the model parameters of $r$, $n_m$ and $n_e$ are set as: $r$ = 2.5 μm, $n_m$ = 1.5 and $n_e$ = 1.0. Due to the axial symmetry of the microcylinder, the incident light with incidence angles ranging from -90° to 0° is considered. The yellow solid line and the purple solid line in Figure 1d show the slope variation of the emergent rays and the position change of the emitting-points along the direction of x-axis. There is a jump point, with $\theta_{\text{iem}\pm}$ = -86.72°, $S_{oe}$ < 0 and $L_x$ > 0, on the slope cure, which divides the slope values into positive and negative parts. From the position distribution curve $L_x$, we can get the dividing point, with $\theta_{\text{iemRW}}$ = -82.82°, $S_{oe}$ > 0 and $L_x$ > 0, between RWC and RRC. The range of RWC is -90° to -82.82°. Because the negative part of the slope curve is at RWC and the range (3.28°) is very small, we can just think about the positive curve where $\theta_{\text{iem}}$ is greater than $\theta_{\text{iem}\pm}$. Through the slope curve, obviously, the curve is changing from steep to flat versus the light from the edge to the middle of the first LRI, and the rate of change of the slope relative to the incidence angle is getting smaller along this process. This explains why the convergence and divergence of light in RRC is faster than that in RSC.

Natural logarithms are often used in the mathematics and physics domain to solve the problems related to growth. Even the exact rate of growth is not known, the cause of the change in growth can be found by taking the log of the function.[47, 48] Figure 1e is the natural log of the positive part of slope function in Figure 1d. The scope includes part of the RWC, the whole RRC and the whole RSC with $\theta_{\text{iem}\pm}$ as the boundary. This is a continuous plane algebraic curve which the outstanding feature is that the curve consists of an upward concave part (magenta-pink solid curve) and a downward convex part (cyan-blue solid curve). According to the concavity and convexity of curve in calculus, the value of the inflection point, marked with green dot in Figure 1e, can be obtained by finding the value of incidence angle ($\theta_{iem}$) where the second derivative of $\ln(S_{oe})$ function is equal to 0.[49, 50] Figure 1f shows the value distributions of $d^2[\ln(S_{oe})]/d\theta_{iem}^2$. This is a nonlinear curve that goes from positive to negative with $\theta_{iem}$ increasing from -86.72° to 0°. It is quite clear that the position of inflection point is in the region of RRC and RSC, which is also the position of boundary point of RRC and RSC. As shown from the inset at the top panel of Figure 1f, the value of the incidence angle at the inflection point is $\theta_{iem}$ = -44.35°.

By using the position of the inflection point obtained, we can precisely delimit the regions of RRC and RSC on the first LRI and execute researches based on this delimitation. Figure 2 shows the comparison of optical field distribution and key parameters of the light-focusing generated from RSC, RRC and the composite region (RRC+RWC) exposed to light alone under the conditions of $\lambda_0$ = 400 nm, $r$ = 2.5 μm, $n_m$ = 1.5 and $n_e$ = 1.0. From the previous calculations, the ranges of RSC, RRC and RWC represented by the interval of incidence angles ($\theta_{iem}$) are [-44.35°, 44.35°], [-82.82°, -44.35°) ∪ (44.35°, 82.82°] and [-90°, -82.82°) ∪ (82.82°, 90°], respectively. Referring to refs. [29, 30], the key parameters of focal distance $f$ (or working distance $W_d$) and beam width ($\omega$) are defined as the distance between the center of the microcylinder (the vertex of the rear surface of the microcylinder) and the $I_{max}$ of the focal spot, and the transverse width of the focal spot using full width at half-maximum (FWHM), respectively. When the incident plane wave illuminates the RSC, as shown in Figure 2a-1, a light-focusing with a profile conforming to the conventional PNJ characteristics is formed on the shadow-side surface of the microcylinder. The focused light field contains only one central main lobe. From the normalized spatial distribution of PNJ in the range of $z = r$ to $z = 3r$ on the $z$-axis (cyan-blue solid line), the focal distance is $f = 3.277$ μm, which is 8 times greater than the illumination wavelength $\lambda_0$. The working distance is $W_d = f - r \approx 1.44\lambda_0$. The normalized transverse light field from $x = -r, z = f$ to $x = r, z = f$ is drawn in Figure 2b-1. A beam width of $\omega = 247.27$ nm is obtained by fitting the distribution curve with a Gaussian function. The corresponding distributions of Poynting vector fields and energy flow diagram represented by black conical arrows and purple solid curves in Figure 2c-1 indicate that the light beams within RSC propagate through the incoming and exiting interfaces with relatively small deflection angles, which results in the formation of a PNJ with long $f$ ($W_d$) and wide $\omega$. When the RRC is illuminated, as shown in Figure 2a-2, a distinct focused light field with $f = 2.639$ μm and $W_d = 139$ nm is formed close to the rear surface of the microcylinder. Such tight focusing consists of a central main lobe and two side secondary lobes. The spacings between the main lobe and the secondary lobes are clear and discernible. As can be seen from Figure 2b-2, the peak intensity of the secondary lobe is slightly more than half of the $I_{max}$. The beam widths of the main lobe, the left-side lobe and the right-side lobe (relative to the frame of reference $z$-axis) are $\omega_c = 106.71$ nm, $\omega_{ls} = \omega_{rs} = 109.50$ nm by using multi-peaks Gaussian fitting, which are far less than the classical diffraction limit (200 nm). The larger deflections of Poynting vector arrows and energy flow streamlines can be clearly seen at the front and rear interfaces of the microcylinder in Figure 2c-2, which leads to the production of specially distributed near-field focusing with short $f$ ($W_d$) and narrow $\omega$. Figures 2a-3, 2b-3 and 2c-3 are, respectively, the light field distribution, transverse FWHM at the focal point along $x$ direction, and Poynting vector distributions and energy power flow for the case of the composite region composed by RWC and RRC illuminated by the plane wave. In comparison with the case where only the RRC is irradiated, the optical field profile and the focal distance $f$ (working distance $W_d$) are almost the same. The differences of the central main lobe and the side secondary lobes are 0.25% and 0.81% for these two cases. It is rarely to see any difference of field and streamline distributions between Figure 2c-3 and Figure 2c-2. The subtle difference in distribution is only found in the two secondary lobes (see Figure S1 in the Supporting Information). By comparing the positions of eleven streamlines on both sides of the $z$-axis for the cases of 'RRC' and 'RWC+RRC', the deviations of the central main lobe within the effective width in the $x$ direction is zero, while the deviations of the side secondary lobes within its effective width is less than 1.47%. Obviously, the impact of RWC on light-focusing is very weak, which is consistent with the previous conclusions upon ray optics analysis. So, we will use the name of composite-RRC (C-RRC) for the region of 'RWC+RRC' in the remaining of this paper. Besides, as the spot intensity of PNJ in the case of RSC irradiated is higher than the spot intensity of near-field focusing in the case of C-RRC irradiated, the focused spot formed by the combined superposition effect of which equivalent to the whole first LRI of the microcylinder irradiated is the PNJ.

It is known from the previous formulas and functions that the position of the inflection point is determined by the *RIC* of dielectric microcylinder and environment medium. The independent variable adopted here is the angle of incidence, which can also be simply substituted by the $x$-coordinate of the intersection between the incident light ray and the first LRI of microcylinder with formula of $|x| = r \cdot \cos \theta_{\text{iem}}$. Figure 3a shows the position of the inflection point along with the *RIC* increasing from 1.35 to 1.65 with step-size of 0.05. The size of the microcylinder is fixed at $r = 2.5$ μm. As can be seen from the chart, the position of the inflection point moves towards the symmetry axis of the microcylinder in an approximate linear way as the *RIC* increases. The linear fitting function calculated by the calculation processing is $\theta_{\text{iem}} \approx 18RIC - 71.43$. The reason for this trend is that the greater the *RIC*, the faster the deflection rate of the microcylinder to the incident light, and thus the earlier the transformation from RSC to RRC. It should be noted that the RWC does not exist for all *RIC*. There is a critical value of $\sqrt{2}$ for *RIC*, where the grazing incidence at $\theta_{\text{iem}} = \pm 90°$ transmits to the right-most vertex of microcylinder and emits.[45] When $RIC > \sqrt{2}$, e.g. $RIC = 1.60$ as shown in the top left panel of Figure 3b, the emergent ray propagates outward away from the symmetry axis of the microcylinder with an emitting-point above the *z*-axis in case of the incident point (*A*) of the plane wave within the region below the blue horizontal line and above the yellow line, while the emitting-points locate below the *z*-axis and propagates towards *z*-axis for the vast majority of the incident light above the blue line and below the *z*-axis. In the graph of $L_x$ with respect to $\theta_{\text{iem}}$, we can clearly see such characteristic for the position distribution of the emitting-points (blue curve). This is similar to the research model described above, in which the contribution to the focused light on the *z*-axis is weak. When *RIC* is equal to the critical value of $\sqrt{2}$ (the top center panel of Figure 3b), with the exception of the incident point *A* right at the lower vertex position of the microcylinder, the emitting-points of the emergent rays are below the *z*-axis, the green curve in the figure, for all the other incident light. When *RIC* decreases below $\sqrt{2}$, e.g. $RIC = 1.35$ in the top right panel of Figure 3b, none of the incoming rays, after transmission through the microcylinder, are going to exit with the positions of the emitting-points on or above the *z*-axis (red curve in the figure). All light emitted from the points *L* with negative *x*-coordinates and propagating towards the *z*-axis is a component of the focal spot on the *z*-axis. When the absence of RWC is understood as the value of RWC is null, we can still use C-RRC to represent 'RWC+RRC' in the following content.

Numerical simulation experiments are carried out to investigate the impact of different *RIC* on the key parameters of PNJ and near-field focusing. Figures 3c and 3d show the numerical results with the same parameters and variables as Figure 3a. Because $W_d$ and $f$ have the same and synchronized trends, we only choose $W_d$ for demonstration. Whether the light shines on the RSC or the C-RRC, the working distance $W_d$ and the transverse beam width of the central main lobe $\omega_c$ are decreas with the increment of *RIC*. For the case of plane wave illuminating the RSC, all the $W_d$ exceeds one wavelength which can be considered that the focused light is the conventionally defined PNJs, and all the beam width are above the diffraction limit. For the case of plane wave illuminating the C-RRC, the characteristics and rules of the focused light produced are more complicated. The working distances of the central main lobe are within a wavelength range except *RIC* = 1.35 ($W_d \approx 1.18\lambda_0$). When *RIC* = 1.65, the focal point returns to the surface of the microcylinder. The beam widths, both the central main lobe shown in Figure 3d or the side secondary lobes not shown here, are far below the diffraction limit. The minimum $\omega_c$ for *RIC* = 1.6, with $W_d$ beyond evanescent decay length ($\sim\lambda/2\pi$),[30] is ~104.34 nm, which may offer promising prospects for nanoimaging, microprobes, micro/nano-manipulation, etc.

By virtue of ray optics approach, the variation trend of $W_d$ and $\omega_c$ can be well analyzed and explained. The slope curve of the emergent light ($S_{oe}$) and the position distribution curve of the emitting-point ($L_x$) with respect to the *x*-coordinate of the incident point ($A_x$) are illustrated in Figures 3e and 3f. Actually, the two set of curves in Figure 3e are coming from a continuous set of curves in the range of (-2.5 μm, 0 μm). For the purpose of making a distinction display of the investigated RSC and RRC, and showing the deviation of the slope curve along the vertical axis with different *RIC* values, we segment each constituent curve at the slope value corresponding to the inflection point of each *RIC* value. The position of the inflection point, in Figure 3e, is the right-most point of each curve in the left set of curves, as well as the left-most point of each curve in the right set of curves. Meanwhile, for a better comparison, we only show the part where the slope value is greater than 0. Based on the previous conclusion of critical value ($\sqrt{2}$), the set of curves on the left in Figure 3e, $A_x$ can cover all values ranging from 0 to -2.5 μm only for the cases of *RIC* = 1.35 and *RIC* = 1.40. The slope values related to the incident points outside the dividing points for the curves corresponding to $RIC > \sqrt{2}$ are not drawn. It can be seen from the slope curves, either associated with RSC or RRC, that the greater the *RIC*, the steeper the slope curves of the emergent rays. The steepness of the slope curve determines how fast the light converges and diverges at the rear surface of the microcylinder.[29] Furthermore, the width of the position distribution for both cases of 'RSC' and 'RRC', as shown in Figure 3f, becomes smaller and smaller with the increases of *RIC*. Taken together, a steeper slope change of emergent ray and a narrower position distribution of emitting-point produce a PNJ or a near-field focusing of shorter working distance $W_d$ and narrower transverse beam width $\omega$; conversely, a smoother slope change of emergent ray and a wider position distribution emitting-points generate a light-focusing of larger $W_d$ and $\omega$.

Based on the study of the inflection point discovered, we then propose and demonstrate a unique way for the modulation of light-focusing generated from plane-wave-illuminated microparticles. In contrast to the traditional way of realizing modulation by using engineering techniques, such as index engineering (refractive index control,[52] *RIC* variation,[29] liquid crystal arrangement,[53] material phase transition,[54] etc.), geometry engineering (multilayer structure,[55] non-spherical structure,[56] engineered structure,[35] etc.) and non-conventional designs,[32] the modulation of light-focusing by our approach can be achieved, without any changes in material attributes and geometrical morphologies of the microparticles, by setting and adjusting the different proportions of electric field intensity illuminating on RSC and RRC. In addition, we can also obtain the generation and modulation of PH in this way, with the approach different from the methods reported so far.[38, 57-59] Figures 4a-1 and 4a-2 show two schematic diagrams of PNJ and PH existing on the shadow-side surface of the microcylinder for two different cases. Taking the inflection points on the first LRI as the cut-off points, the electric field intensities illuminating the C-RRC above, the RSC and the C-RRC below are set as $|E_u|$, $|E_m|$ and $|E_l|$, respectively. The subscripts u, m and l denote the electric field applied to the upper, middle and lower regions of the microcylinder. The size and refractive indices of the microcylinder and surrounding medium are the same as those in Figure 2. Case 1 and 2 represent the illumination conditions of $|E| = |E_u| = |E_l|$, and $|E| = |E_u|$, $|E_l| \neq |E_u|$, where $|E|$ is set as the independent variable of the study. The electric field $|E_m|$ and the illumination wavelength for each region ($\lambda_u = \lambda_l = \lambda_m = \lambda_0 = 400$ nm) are the same for case 1 and 2. As the edge region above and below are symmetric relative to the *z*-axis, the PHs in both the upper and lower directions indicated by red solid spot and light-blue dotted line in Figure 4a-2 will be formed when the electric field on one side is modulated.

Figure 4b illustrates the modulation results of PNJ key parameters for case 1. The yellow and purple axes are the functions of $W_d$ and $\omega_c$ with respect to $|E|/|E_m|$, respectively.

During the modulation process, the intensities of $|E_u|$ and $|E_l|$ are always the same and change synchronously. The abscissa variable $|E|/|E_m|$ is from 0 to ∞, where the leftmost 'start' symbol for $|E|/|E_m| = 0$ represents no incident plane wave illuminating the RSC and the rightmost 'stop' symbol for $|E|/|E_m| \to \infty$ means infinitely close to the case of the C-RRC above and C-RRC below irradiated by a plane wave. The interval between 0.1 and 500 is divided into four segments, where the step-sizes set in the intervals of [0.1, 0.5], [1, 5], [10, 50] and [100, 500] are 0.05, 0.5, 5 and 50, respectively. With the increase of $|E|/|E_m|$, both $W_d$ and $\omega_c$ show a nonlinear downward trend. The average rate of change (ARC), calculated by dividing the difference of $W_d$ by the difference of $|E|/|E_m|$ for the two endpoints in each interval, is used to characterize the trend of nonlinear change. The ARCs in the above four intervals, marked in Figure 4b, are -280.25 nm, -38.125 nm, -1.46 nm, -0.02 nm for $W_d$ vs $|E|/|E_m|$ function and -91 nm, -13.38 nm, -0.13 nm, -8.5×10$^{-4}$ nm for $\omega_c$ vs. $|E|/|E_m|$ function. It is observed that the decline of $W_d$ and $\omega_c$ probably goes through three stages of fast, slow and steady. $\omega_c$ reaches its stable value faster than $W_d$. The stable value of $\omega_c$ is almost exactly the same as the $\omega_c$ for the case of only the C-RRC illuminated by the light. This means that there is no need to infinitely increase the $|E|/|E_m|$ for modulating the key parameters of the light-focusing. In fact, the narrowing of $\omega_c$ is less than 5% after $|E|/|E_m| > 10$ (~3.07% for $|E|/|E_m|$ between 15 and 500). The insets of Figure 4b show the contour plots of light field distributions for $|E|/|E_m|$ equaling 0.3, 1, 3, 30 and 300. When $|E|/|E_m| < 1$, the resulting PNJ has only one central main lobe. When $|E|/|E_m| \geq 1$, the side secondary lobes appear and become more and more apparent. When $|E|/|E_m|$ is greater than a certain value, e.g. $|E|/|E_m| = 30$ and $|E|/|E_m| = 300$, the contour distributions of the focusing constituted by main lobe and secondary lobes change very little.

Case 2 describes the situation where $|E_l|$ and $|E_u|$ are not equal and out of sync. Two examples of $|E_u|/|E_m| = 0$, $|E_l|/|E_m| = 1.5$ and $|E_u|/|E_m| = 3.5$, $|E_l|/|E_m| = 0.5$ are shown in Figures 4c-1 and 4c-2, respectively. $|E_m|$ equals unit intensity. The optical field distributions displayed by heat-camera-color-filled contour plots draw the outline of two typical PHs with bending angles in different directions. The positions of the maximum light intensity $I_{max}$ (white dot marked in Figure 4c) are at $x = 2.98$ nm, $z = 3.025$ μm for Figure 4c-1 and $x = 17.56$ nm, $z = 2.884$ μm for Figure 4c-2, respectively. With reference to our previous work,[58] two trajectory lines connected by several small linear segments, marked as Cambridge-blue and wine-red dotted lines in the two figures, show the optical field distribution along two curved trajectories. Figure 4c-3 is a 3D display diagram, showing the light intensity of the points on the two trajectories, in the $xOz$ plane. The features of bending distribution are clearly visible as well. The normalized transverse optical field distribution at the $I_{max}$ point along $x$ direction for these two samples are shown in Figure 4c-4. The FWHMs obtained by Gaussian function fitting are ~$0.51\lambda_0$ and ~$0.44\lambda_0$, respectively. Obviously, the higher the intensity of light illuminating the edge region of the microcylinder, the more efficiency the C-RRC in converging and diverging light, and hence the smaller the $W_d$ and $\omega_c$ of the produced PH. As can be seen from the results presented in this part, our method is different from the PH generation with common ways of breaking the symmetry of the microparticles in geometrical structure[57] and material composition.[58] The PHs generated through these two common ways are mainly from the interference of light with different phase velocities, and the differences in the phase velocity are induced by the different structure or material of microparticle's upper and lower parts. Our method proves that PH can be produced by only setting the intensity of the light irradiating to the

C-RRC above and C-RRC below to be different. It also shows that the PH phenomenon does not mean light travelling in curved lines, instead, the light of different states interferes with each other to form a curved distribution of light field.

Therefore, the PHs with different profiles can be easily created and effectively modulated by adjusting the values of $|E_u|/|E_m|$ and $|E_l|/|E_m|$. Figure 4d-1 shows the simulation results of working distance $W_d$ and transverse beam width $\omega_c$ depending on $|E_u|/|E_m|$ for the cases of $|E_l|/|E_m| = 0$ and $|E_l|/|E_m| = 0.3$. As can be seen from the top left and bottom left panels of the figure, the value of $W_d$ decreases as $|E_u|/|E_m|$ increases, while the value of $\omega_c$ first decreases and then increases. The minimum $\omega_c$ is ~$0.49\lambda_0$ at $|E_u|/|E_m| = 2.25$. The reason why $\omega_c$ goes down in the first place is because, as $|E_u|/|E_m|$ goes up, the edge part representing the light with a faster ability to converge and diverge plays a bigger and bigger role. Consequently, under the conditions of the focal spot intensity arising from the C-RRC illuminated not much more than the focal spot intensity from the RSC illuminated, the combined effect of these two focusing causes the focal point getting closer and closer to the microcylinder, and the beam waist of the focal spot becoming smaller and smaller. $\omega_c$ is getting bigger, when $|E_u|/|E_m|$ reaches 2.5 and above, because the light-focusing related to C-RRC is starting to dominate. In the case of boundary illumination (equivalent to the C-RRC illuminated in this paper), the $\omega_c$ and the distance off the $z$-axis for each transverse optical field, in the region from the focal point to the surface of the microcylinder, increases as the distance between transverse field and microcylinder surface decreases.[44] As $|E_u|/|E_m|$ goes from 2.75 to 3, there has been a significant change, that is, a jump in $W_d$ and $\omega_c$. When the case of $|E_l|/|E_m| = 0$ is viewed as the superposition of RSC irradiated and C-RRC above irradiated, the distribution of light intensity on the curved

trajectory has two peaks, which, similar to the wine-red marked curve, 3D curve and solid curve in Figures 4c-2, 4c-3 and 4c-4, respectively. Figure S2 clearly shows where these two peaks are located in the *xOz* plane for $|E_u|/|E_m| = 2.5$, $|E_u|/|E_m| = 2.75$, $|E_u|/|E_m| = 3$ and $|E_u|/|E_m| = 3.25$, and the jump in $I_{max}$, respectively. The $W_d$ and $\omega_c$ for the case of $|E_l|/|E_m| = 0.3$, as shown in the right panels of Figure 4d-1, have experienced similar rule of change. Just because the C-RRC below is also illuminated by the plane wave, when the values of $W_d$ and $\omega_c$ become less and less affected by the increasing light exposure to C-RRC above, the focused field generated from plane-wave-illuminated C-RRC below will cancel out and decelerate this reduction. Thus, the $W_d$ and $\omega_c$ for each $|E_u|/|E_m|$ value in case of $|E_l|/|E_m| = 0.3$ are smaller than the case of $|E_l|/|E_m| = 0$. The jump also occurs when $|E_u|/|E_m|$ is larger. There is no further discussion after the jump, because the $W_d$ (~25.8 nm) is already less than the evanescent decay length (~$\lambda_0/2\pi \approx 64$ nm). Similarly, the value of $W_d$ (~59.7 nm) is less than this decay length at $|E_u|/|E_m| = 5.5$ for case $|E_l|/|E_m| = 0$ as well. And for $W_d$ in both cases, the value is always going to smaller as the focus is always moving towards the surface of the microcylinder.

Bending angle ($\delta_b$) is a particular characteristic parameter of PH distinct from PJ.[57, 58] At present, there are many definitions of bending angles, which are basically defined around the $I_{max}$ point, start point and end point.[39, 40, 42, 57-61] $I_{max}$ point is also often referred to as the 'bend point' and 'inflection point'.[42, 57] Of course, the 'inflection point' here is not the same as the inflection point discussed in this paper, but refers to the deflection of the PH beam around this point. In fact, the point at which the deflection occurs cannot be named as the 'inflection point', because there are several points where the PH beam deflects in some cases. We will give a detailed example in the Supporting Information,

Section 4. So, we named this point '$I_{max}$ point' in this article due to the uniqueness of '$I_{max}$ point' in PH. As for the start point and end point, there are always some problems such as fuzzy definition, unclear description, no or unclear explanation, etc. Through the reference and synthesis of these methods, we propose a method which can accurately define and calculate the bending angle. The method uses the points with light intensity of $I_{bmax}$, $I_{max}$ and the farthest point on the $I_{max}/e^2$ contour curve, which each of these three points has a unique and definite position, as the start point, $I_{max}$ point and end point. The detailed graphical presentation and text description are available in the Supporting Information, Section 3. It is important to note that the bending angle is just one way of describing the curvature characteristic of PH. This is an artificial parameter, not an intrinsic parameter of PH. This is not to say that there can be only one definition of it. In order to quantify the bending angle, we give a new definition here to uniquely and certainly calculate the value of $\delta_b$.

Figure 4d-2 shows the variation of bending angle $\delta_b$ in both cases as shown in Figure 4d-1. As in ref. [58], the angles of deflection in the clockwise and counterclockwise directions are set as negative and positive, respectively. From the previous analysis, the optical field of the main lobe is biased to the C-RRC side with loaded illumination light. The $I_{bmax}$ point is above the z-axis and $I_{max}$ point, and all the bending angles are negative for the case of $|E_l|/|E_m| = 0$. In the process of $|E_u|/|E_m|$ going from 0 to 0.75, the banding angles change very quickly from 0° to the minimum value of -24.09°. When $1 \leq |E_u|/|E_m| \leq 5$, the value of $\delta_b$ is around -20°, and the deviation between the maximum and minimum value is 6.11°. For $|E_l|/|E_m| = 0.3$, the $\delta_b$ is positive in the range of $0 \leq |E_u|/|E_m| < 0.3$. In the negative part of $\delta_b$, the change rate of $\delta_b$ is slower than the case of $|E_l|/|E_m| = 0$ due to the cancellation effect induced by the light illumination of C-RRC below. Because of this cancellation effect, the $I_{max}$ point overall is closer to the central axis which, in general, the $\delta_b$ for $|E_l|/|E_m| = 0.3$ is less than the $\delta_b$ in case $|E_l|/|E_m| = 0$. The maximum difference between positive and negative bending angles is ~40.1°. The illustration shows two examples of PHs with $\delta_b = -17°$ for $|E_l|/|E_m| = 0$, $|E_u|/|E_m| = 1.25$, and $\delta_b = -21.07°$ for $|E_l|/|E_m| = 0.3$, $|E_u|/|E_m| = 2.25$. The dotted circle marked cases are the generated PNJs in the condition of $|E_u|/|E_m| = |E_l|/|E_m|$.

## Conclusions

In this paper, starting from the third rule of light focusing produced by plane-wave-illuminated microparticle, the inflection point and the PNJ phenomenon on the inflection point perspective have been thoroughly discussed. The inflection point is a special point on the first LRI of the microparticle, which can accurately delimit and divide the regions where the light converges and diverges with rapid or slow modes, i.e. the RRC or the RSC. The position of the inflection point can be determined by the sharpness of light-focusing reflected by the slope curve of emergent rays, and the practical significance of natural logarithm function and the second derivative of a function. Based on this partition and the other partition RWC, PNJ with only one central main lobe and near-field focusing with one central main lobe and two side secondary lobes can be obtained by separately illuminating the RSC, RRC and 'RRC+RWC' of the microcylinder, respectively. The key parameters, i.e. the focal distance $f$ (working distance $W_d$) and the transverse beam width $\omega$ for RSC are long and wide, while for RRC and 'RRC+RWC' are short and narrow. The deviations of focal distance (working distance), beam width of central lobe and side lobes are respective 0%, 0.25% and 0.81% for the later two cases. With the increase of *RIC*, the position of the inflection point approaches the central symmetry axis of the microcylinder in an approximate linear way. RWC only exists if *RIC* is less than the threshold of $\sqrt{2}$. The FEM-based simulation results and ray optics-based theoretical analyses for *RIC* varying from 1.35 to 1.65 jointly verify that the variations on the working distance $W_d$ and the transverse beam width of the central main lobe $\omega_c$ follow a downward trend. In the case that $|E_u|$ and $|E_l|$ are equal and change simultaneously, the produced PNJ can be modulated, in a nonlinear way, by adjusting the field intensity of the illumination light on RSC and the composite RRC. The modulation ranges are over $1.5\lambda_0$ for $W_d$ and 106.44 nm to 247.27 nm for $\omega_c$. In the case of $|E_u| \neq |E_l|$ and out of sync, a new method for the generation of PH can be obtained. When tuning $|E_u|/|E_m|$ from small to large, the key parameters and bending angle ($\delta_b$) of the PH can be flexibly modulated. The modulation range of the characteristic parameter $\delta_b$ for $|E_u|/|E_m|$ ranging from 0 to 5 in the cases of $|E_l|/|E_m| = 0$ and $|E_l|/|E_m| = 0.3$ are 24.09° and 40.1°, respectively. The modulation of PNJ and PH does not require any changes in the morphological structures and material properties of the microparticles. By the through study on the inflection point, our work theoretically solves the problem of dividing RSC and RRC on the first LRI of microparticles in PNJ field, which will enable deeper insights into the nature of light-particle interaction.

## Conflict of Interest

The authors declare no conflict of interest.

## Acknowledgements

This work was supported by the Guangdong Basic and Applied Basic Research Foundation (2019A1515011242), Key-Area Research and Development Program of Guangdong Province (2019B020226004), National Natural Science Foundation of China (61805271, 62074155), Shenzhen Science and Technology Innovation Commission (JCYJ20170818154035069, KCXFZ202002011008124), and CAS Key Laboratory of Health Informatics (2011DP173015).

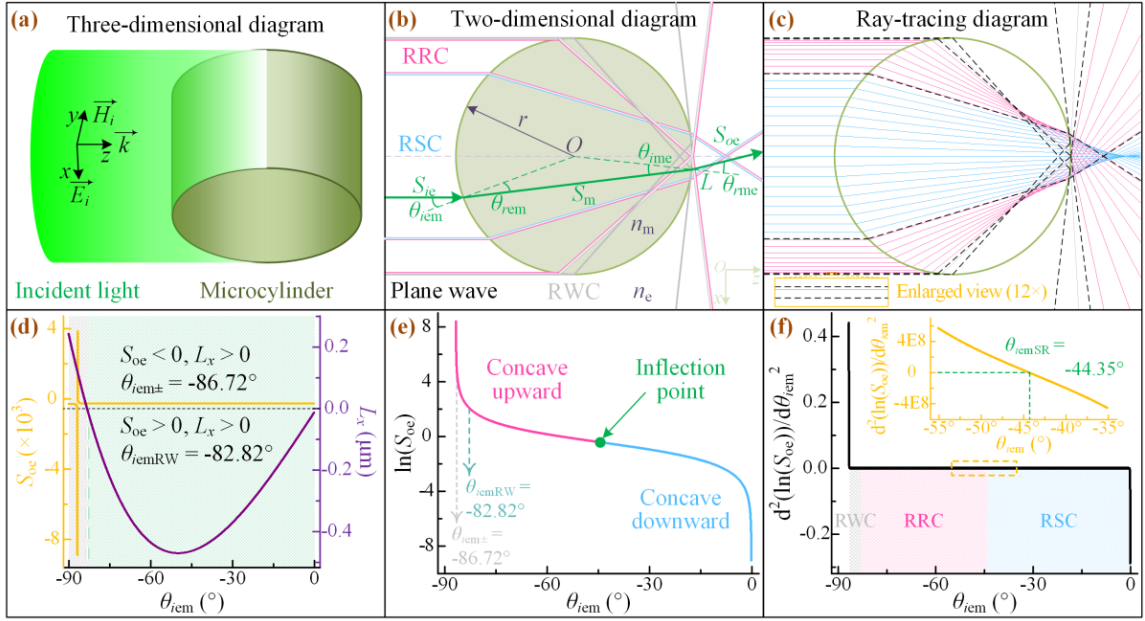

**Figure 1.** Schematic illustration on the research model and the obtained inflection point. a) 3D schematic diagram of a dielectric microcylinder illuminated by an incident plane EM wave. b) 2D representation of the research model in a). Green solid lines show the process of a plane wave propagation through the microcylinder. RSC: region of slow change, RRC: region of rapid change, RWC: region of weak contribution. $O$: origin of coordinates, $r$: radius of microcylinder, $n_m$ & $n_e$: refractive indices of microcylinder and environmental medium. $\theta_{iem}$, $\theta_{ime}$: angle of incidence, $\theta_{rem}$, $\theta_{rme}$: angle of refraction, $S_{ie}$, $S_m$ & $S_{oe}$: slopes of the transmission rays. c) Ray trajectories of the plane light wave transmission through RWC (light-gray solid line), RRC (magenta-pink solid line) and RSC (cyan-blue solid line). Inset: 12× enlarged view of the selected area in RWC. d) The slope of emergent ray ($S_{oe}$) and the position of emitting-point along the $x$-axis ($L_x$) as a function of the incidence angle ($\theta_{iem}$). $\theta_{iem\pm}$: the angle of incidence corresponding to the jump point between positive and negative values of the slope curve, $\theta_{iemRW}$: the angle of incidence corresponding to the dividing point between RWC and RRC. e) The natural log of the slope function $\ln(S_{oe})$ with respect to the initial incidence angles $\theta_{iem}$. f) The second derivative values of the slope curve shown in e). Inset: an enlarged view of $d^2[\ln(S_{oe})]/d\theta_{iem}^2$ in the range of $\theta_{iem}\in$(-55.08°, -35.08°).

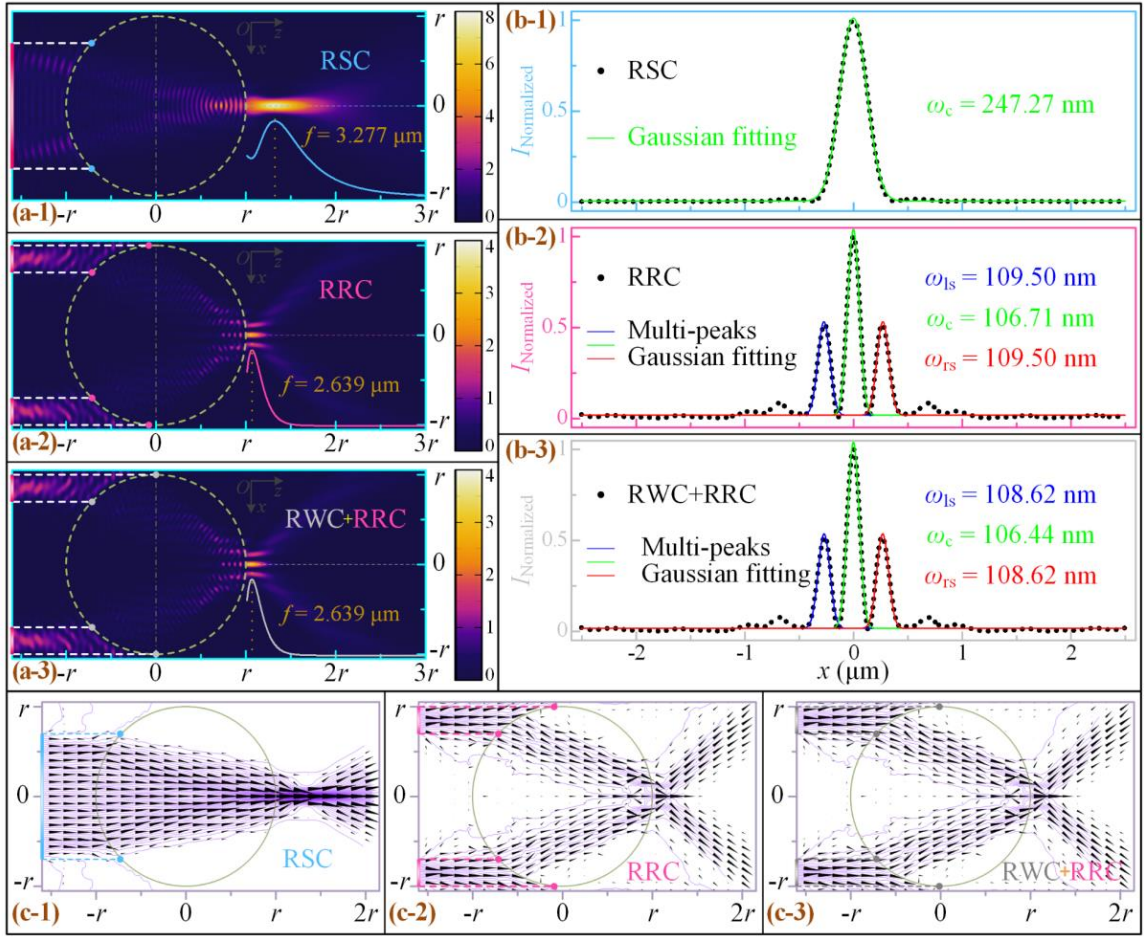

**Figure 2.** 2D optical field distributions of light illuminating the a-1) RSC, a-2) RRC and a-3) composite region of 'RWC+RRC' obtained from FEM-based full-wave simulations. The model parameters are: $\lambda_0$ = 400 nm, $r$ = 2.5 μm, $n_m$ = 1.5, $n_e$ = 1.0. Insets: the normalized spatial intensities of the focused light field from $x = 0$, $z = r$ to $x = 0$, $z = 3r$ for the cases of RSC (cyan-blue solid line), RRC (magenta-pink solid line) and 'RWC+RRC' (light-gray solid line) irradiated by the plane wave. The vertical dotted line indicates the position of the focal point and focal distance $f$ for each figure. b-1), b-2) and b-3): The normalized transverse light field profiles (black dots) at the focal point along $x$ direction and the single-peak, multi-peaks Gaussian fitting curves corresponding to Figures 2a-1, 2a-2 and 2a-3, respectively. c-1), c-2) and c-3): Poynting vector distributions (black conical arrows) and energy flow streamlines (purple solid curves) for the cases of the irradiation areas located in the regions of RSC, RRC and 'RWC+RRC', respectively.

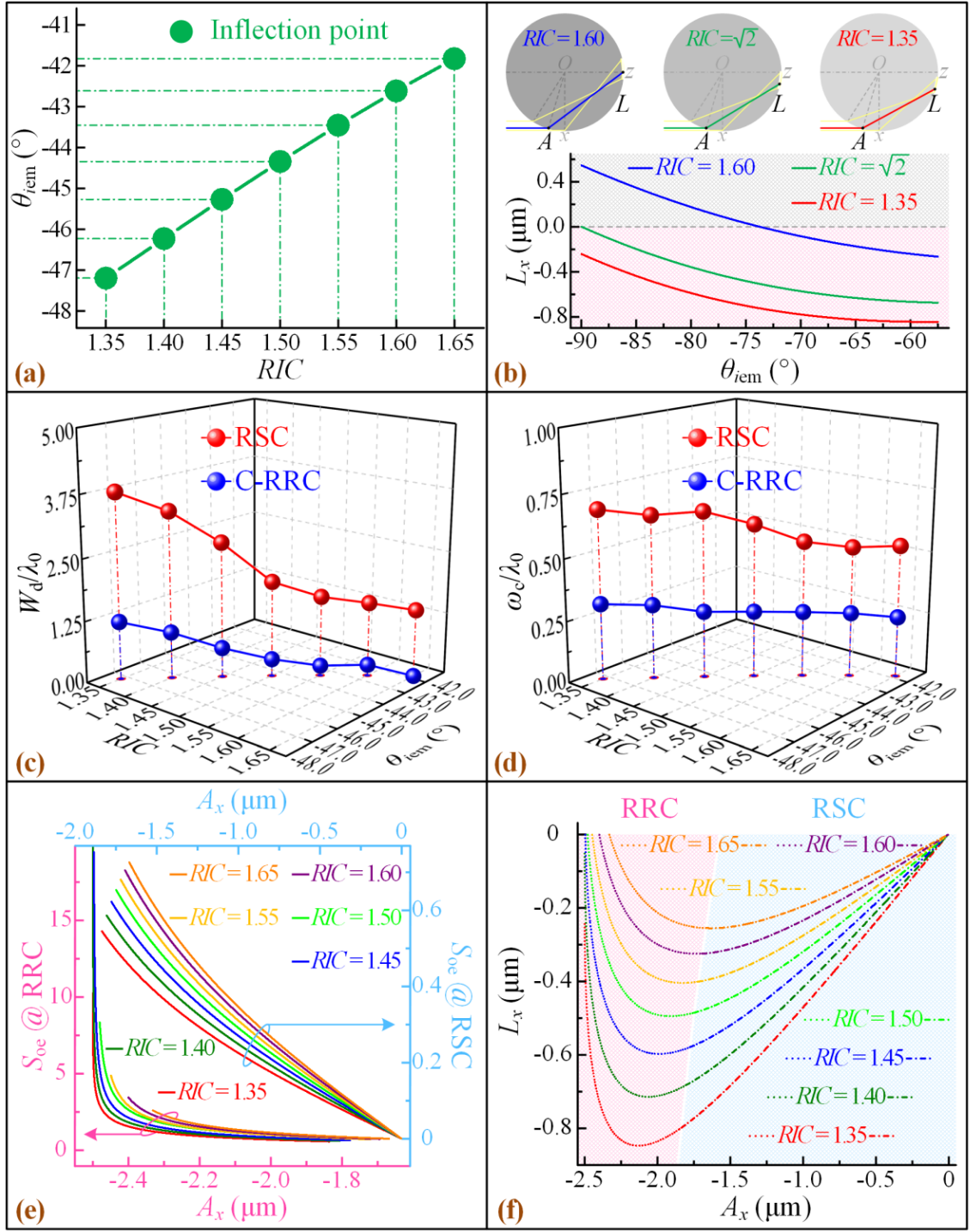

**Figure 3.** a) The inflection point between RSC and RRC of the first LRI for the model of plane-wave-illuminated microcylinder varies with the $RIC$ of microcylinder and environment medium. b) The position distributions of the emitting-points $L$ relative to the incidence angles $\theta_{iem}$ along $x$ direction for the cases of $RIC = 1.60$, $RIC = \sqrt{2}$ and $RIC = 1.35$, respectively. Insets: schematic illustrations of the ray tracing process of propagation light beam and trajectories of the emitting-points for $RIC = 1.60$, $RIC = \sqrt{2}$ and $RIC = 1.35$. $A$: incident point, $L$: emitting point. c) Working distance $W_d$ and d) transverse beam width (the central main lobe) $\omega_c$ of the light-focusing as a function of $RIC$ and the incidence angle at the inflection point associated with the $RIC$ for the cases of RSC and C-RRC illuminated by the plane wave. e) The slope curves of emergent rays $S_{oe}$ and f) the corresponded position distributions of emitting-points $L_x$ as a function of the $x$-coordinate of incident points $A_x$ for different $RIC$. The magenta-pink and cyan-blue coordinate axes in e) and the magenta-pink and cyan-blue shadow areas in f) are relevant to the illumination regions of RRC and RSC, respectively.

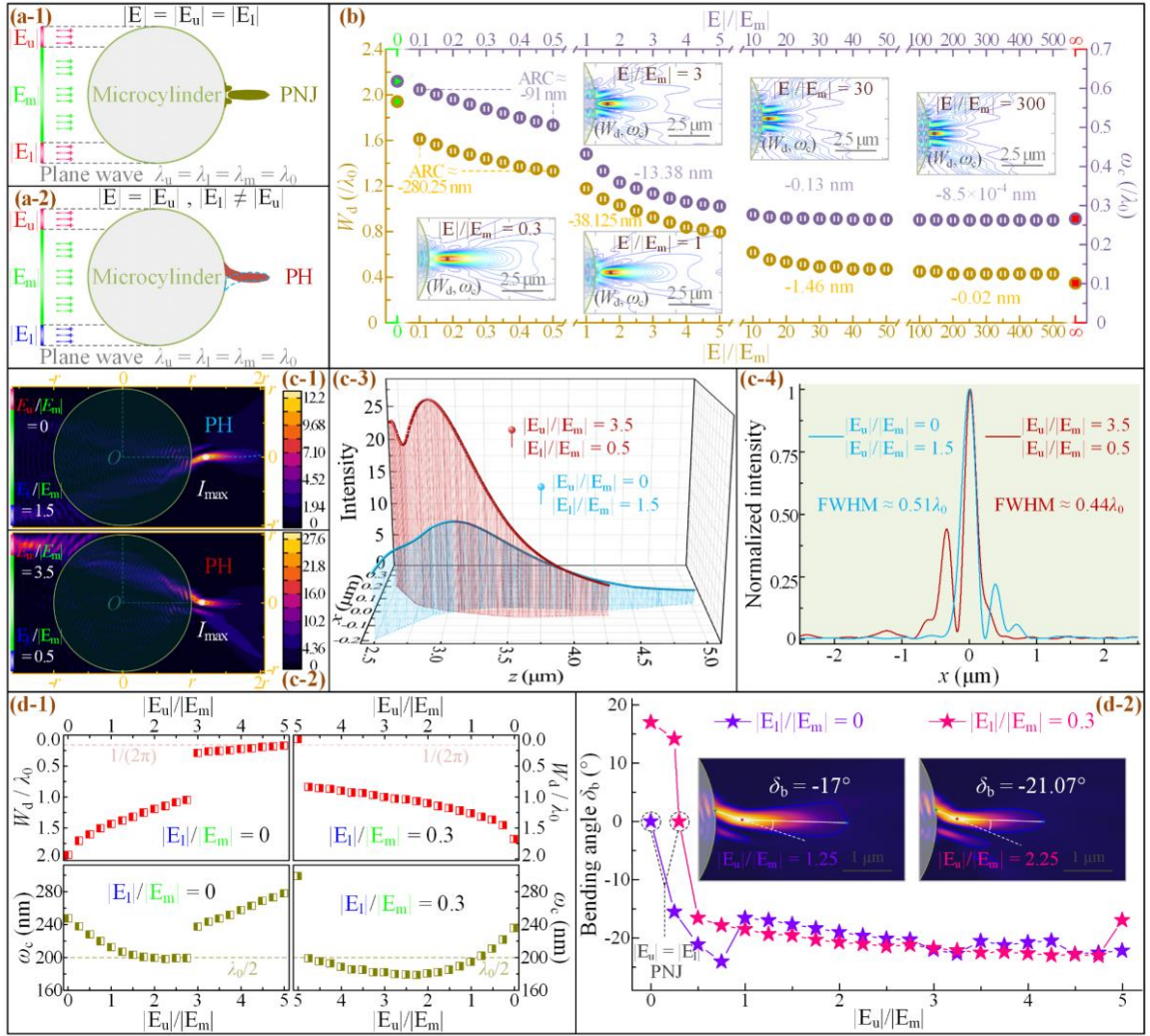

**Figure 4.** Schematic diagrams illustrate the existence of a-1) PNJ and a-2) PH for the cases of $|E| = |E_u| = |E_l|$ and $|E| = |E_u|, |E_l| \neq |E_u|$, respectively. $|E_u|$, $|E_m|$ and $|E_l|$ correspond to the electric field intensity of light irradiated to the C-RRC above, RSC and C-RRC below, respectively. $|E|$ is the independent variable set. $\lambda_u$, $\lambda_m$ and $\lambda_l$ are the wavelengths of the plane waves illuminating each area, where $\lambda_u = \lambda_l = \lambda_m = \lambda_0 = 400$ nm. b) Modulation of PNJ key parameters (working distance $W_d$ and transverse beam width $\omega_c$) by varying $|E|/|E_m|$ from 0 to ∞ for the case of $|E| = |E_u| = |E_l|$. Four intervals of [0.1, 0.5], [1, 5], [10, 50] and [100, 500] with respective step-sizes of 0.05, 0.5, 5 and 50 are contained within the range of the independent variables. ARC: average rate of change stands for dividing the difference of $W_d$ by the difference of $|E|/|E_m|$ for the two endpoints in each interval. Insets: five examples of light field distributions in term of contour lines. The gray dots mark the location of the $I_{max}$. PHs formed at the rear surface of the microcylinder under the illumination condition of c-1) $|E_u|/|E_m| = 0$, $|E_m| = 1$, $|E_l|/|E_m| = 1.5$ and c-2) $|E_u|/|E_m| = 3.5$, $|E_m| = 1$, $|E_l|/|E_m| = 0.5$. c-3) Optical field

distribution of the trajectories marked with Cambridge-blue and wine-red dotted curves in c-1) and c-2). c-4) Transverse optical field profiles at the focal point along *x* direction corresponding to the cases of $|E_u|/|E_m| = 0$, $E_l|/|E_m| = 1.5$ (Cambridge-blue curve) and $|E_u|/|E_m| = 3.5$, $|E_l|/|E_m| = 0.5$ (wine-red curve). d-1) Modulation of PH key parameters ($W_d$ and $\omega_c$) by varying $|E_u|/|E_m|$ from 0 to 5 in the cases of $E_l|/|E_m| = 0$ and $E_l|/|E_m| = 0.3$. The light-red and olive dotted lines marked in the figure indicate the vertical positions of evanescent decay length ($1/2\pi$) and diffraction limit ($\lambda_0/2$). d-2) Bending angles ($\delta_b$) with respect to $|E_u|/|E_m|$ from 0 to 5 in the cases of $E_l|/|E_m| = 0$ and $E_l|/|E_m| = 0.3$. Insets: electric field distributions and bending angles for the cases of $|E_l|/|E_m| = 0$, $|E_u|/|E_m| = 1.25$ and $|E_l|/|E_m| = 0.3$, $|E_u|/|E_m| = 2.25$. Five-pointed stars marked with dotted circle: PNJ generated at $|E_u| = |E_l|$.

# SUPPORTING INFORMATION

# Inflection Point: A New Perspective on Photonic Nanojets

*Guoqiang Gu, Pengcheng Zhang, Sihui Chen, Yi Zhang and Hui Yang*[*]

Dr. G. Gu, Dr. P. Zhang, Mr. S. Chen, Dr. Y. Zhang, Dr. H. Yang
Laboratory of Biomedical Microsystems and Nano Devices
Bionic Sensing and Intelligence Center
Institute of Biomedical and Health Engineering
Shenzhen Institutes of Advanced Technology
Chinese Academy of Sciences
Shenzhen 518055, China
E-mail: hui.yang@siat.ac.cn

Dr. H. Yang
CAS Key Laboratory of Health Informatics
Shenzhen Institutes of Advanced Technology
Chinese Academy of Sciences
Shenzhen 518055, China

## 1. Comparison of energy flow streamline distribution

In order to illustrate the subtle difference of the optical field distributions between the cases of plane wave illuminating the region of rapid change (RRC) and composite region (the region of weak contribution (RWC) + RRC) of the first light refraction interface (LRI), the distributions of energy flow streamlines are compared in the enlarged views. As shown in Figures S1a-1 and S1a-2, the eleven streamlines symmetrically distributed on both sides of the $z$-axis and, marked with small circles and Arabic numbers, are used for analysis. The radius and the refractive index of the microcylinder are $r = 2.5$ μm and $n_m = 1.5$, respectively. The surrounding medium is air ($n_e = 1.0$). The $x$ positions along the dashed lines of $z = f = 2.639$ μm for these two cases are the areas of particular interest. The relative difference of the $x$ positions is set as $\frac{\Delta x}{x_{RRC}} \times 100\% = \frac{|x_{RWC+RRC} - x_{RRC}|}{x_{RRC}} \times 100\%$. Figure S1b shows the relative differences of the eleven streamlines between the investigated two cases. Overall, with the exception of the outmost streamline 11 which is slightly more than 4%, all the relative differences are less than 1.47%. In fact, only the light near the central main lobe and the side secondary lobes is effective for the generation of the near-field focusing spot. Here, referring to ref. [S1], the mean transverse half-width is defined as the distance between the point of the maximum light intensity ($I_{max}$) and the point located at the $1/e^2$ level of $I_{max}$. In this case, there is no relative difference for the central main lobe (the region between the two dotted green lines). For the side secondary lobes, only streamline 4 has a small deviation of 1.47%, while the other streamlines have no relative differences between the cases of RRC and 'RWC+RRC' within the effective transverse widths of the main/side lobes (the regions between two dotted blue or red lines).

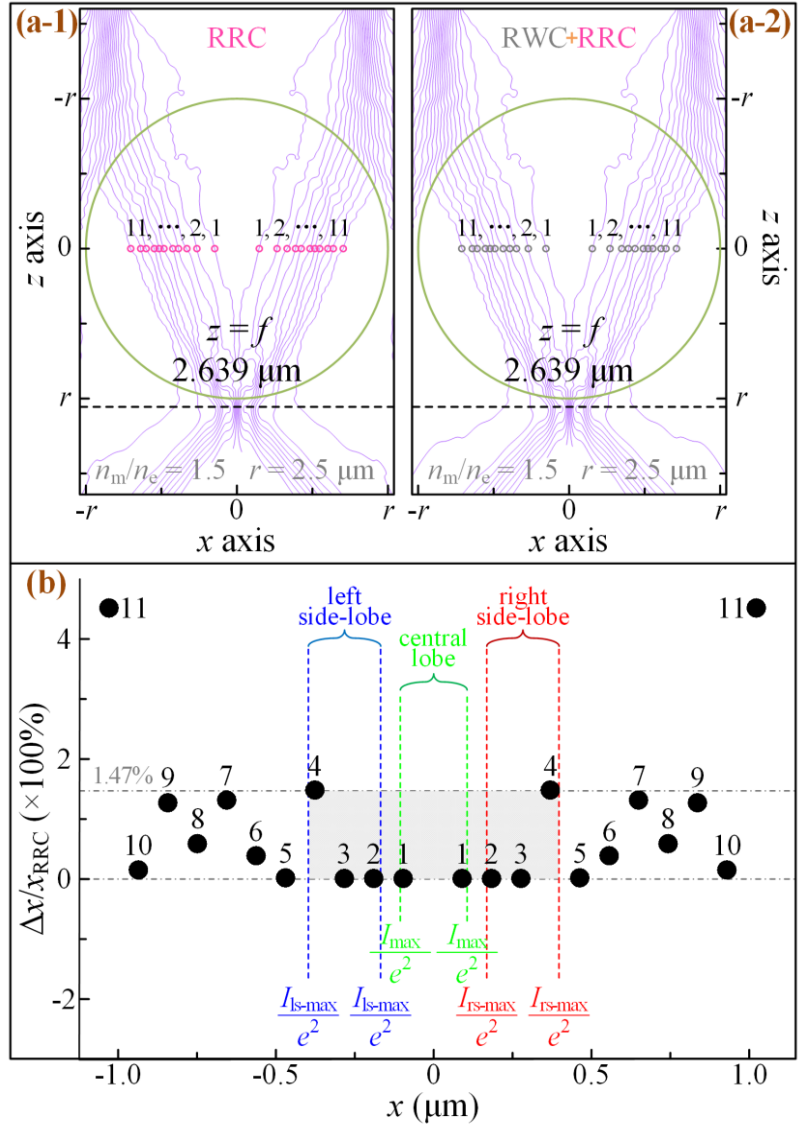

**Figure S1.** Distributions of energy flow streamlines of a-1) RRC and a-2) 'RWC+RRC' on the first LRI that is illuminated by a plane wave. From the central symmetry axis of the microcylinder to both sides, the eleven streamlines are represented by the Arabic numbers 1-11 and marked with small circles. b) Relative differences in $x$ positions of the eleven streamlines for the two cases. The effective widths are the distance between two points of intensity $I$ equaling $I_{max}/e^2$ for the central lobe, and between two points of $I = I_{ls\text{-}max}/e^2$ on the left and $I = I_{rs\text{-}max}/e^2$ on right for the left and right side-lobe, respectively.

## 2. Jump in working distance $W_d$ and transverse beam width $\omega_c$

The research results in Figure 4c tell us that, when the composite-RRC (C-RRC) above and C-RRC below on the first LRI of the microcylinder are exposed to different intensity of light, the distribution of light intensity along the trajectory of the formed photonic hook (PH) has two peaks. The two peaks locate on the rear surface of the microcylinder at the point of $I_{max}$ and the point of local maximum light intensity ($I_l$) on the trajectory line. Point $I_{max}$ is always in a continuous space enclosed by a contour curve. Figures S2a-d are the contour map representations of PHs obtained by simulation under four conditions of $|E_u|/|E_m| = 2.5, 2.75, 3$ and $3.25$, respectively. The parameters of $|E_m| = 1$ and $|E_l|/|E_m| = 0$ are the same for each case. When $|E_u|/|E_m| = 2.5$, the position of $I_{max}$ ($z = 2.936$ μm) is to the right of $I_l$ ($z = 2.636$ μm). The difference of light intensity between $I_{max}$ and $I_l$ is 0.48. When $|E_u|/|E_m| = 2.75$, the intensity difference decreases to 0.07. Neither $I_{max}$ nor $I_l$ has changed much for their positions relative to the case of $|E_u|/|E_m| = 2.5$. At this time, a closed contour curve, marked within the black solid circle, around $I_l$ point is formed, which explicitly indicates a local maximum existing within this enclosed interval.[S2] By comparing Figures S2b and S2c, we can find that the position of $I_{max}$ and $I_l$ are almost exactly switched between $|E_u|/|E_m| = 2.75$ and $|E_u|/|E_m| = 3$. This is exactly the position where $W_d$ and $\omega_c$ jump in the text of the article. The contour plots of the optical field distribution and the calculated light intensity of the two peaks clearly show this jump. The continuous increase of $|E_u|/|E_m|$ further strengthens the dominant effect of light-focusing from plane-wave-illuminated C-RRC region. Therefore, on the whole, the positions of the two peaks are close to the surface of the microcylinder.

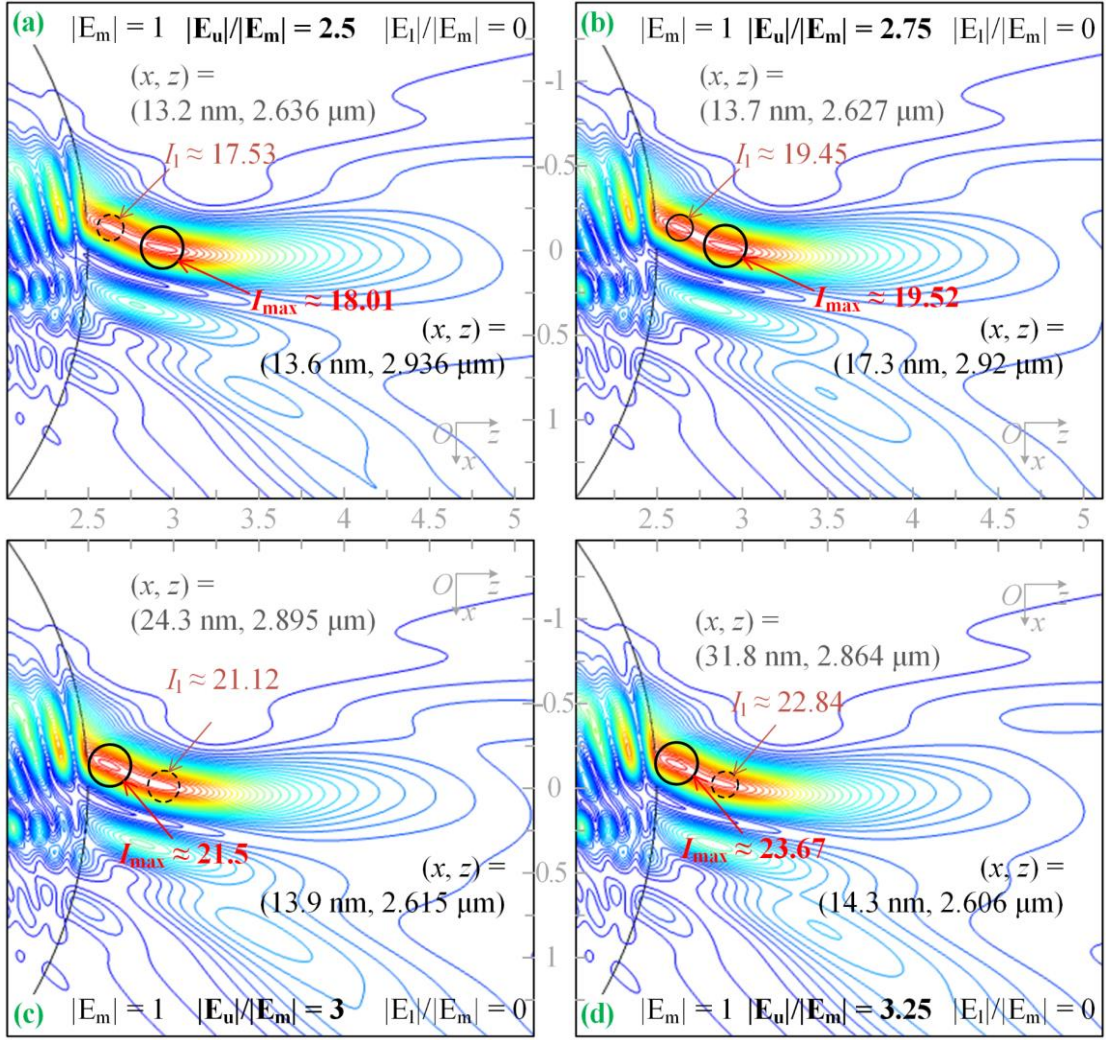

**Figure S2.** Contour plots of optical field distributions for the illumination conditions of a) $|E_m| = 1$, $|E_l|/|E_m| = 0$, $|E_u|/|E_m| = 2.5$, b) $|E_m| = 1$, $|E_l|/|E_m| = 0$, $|E_u|/|E_m| = 2.75$, c) $|E_m| = 1$, $|E_l|/|E_m| = 0$, $|E_u|/|E_m| = 3$ and d) $|E_m| = 1$, $|E_l|/|E_m| = 0$, $|E_u|/|E_m| = 3.25$. $I_{max}$: the maximum light intensity in the entire optical field distribution, $I_l$: the local maximum light intensity within the selected area. The large circles drawn with solid black lines denote a closed interval containing $I_{max}$ point and closed contour curve. The small circles drawn with solid black lines and dotted black lines denote the closed interval containing $I_l$ point and closed contour curve and the unclosed intervals only containing $I_l$ point. The x-coordinate and z-coordinate variables are in microns.

## 3. Definition and description of bending angle $\delta_b$

The key to defining and calculating bending angle $\delta_b$ is to determine the unique and definite positions of the start point, the $I_{max}$ point and the end point. Figure S3a shows the schematic diagram for defining PH's bending angle $\delta_b$. $I_{max}$ point, marked with red dot, is easy to find by numerical calculation as it is the point of maximum light intensity in PH field distribution. The start point is obviously located on the boundary curve, between the microcylinder and the environment medium, with light emitting-points distributed. In terms of decay length of PNJ or PH, the point with light intensity of $I_{max}/e^2$ is selected as the reference for the end point.[S3-S6] A section of boundary curve 'B', marked with green dotted arc line, containing optical field distribution is selected, and the point with the maximum light intensity ($I_{bmax}$) on 'B' is selected as the start point (green dot). The blue curve is a contour line of light intensity equal to $I_{max}/e^2$. In the selected rectangular area 'R' marked in the figure, the right-most point of the contour line $I_{max}/e^2$ is defined as the end point (blue dot). The cyan contour line of $I < I_{max}/e^2$ is used to assist describing the PH profile. Figures S3b-d show an example of calculating PH bending angle $\delta_b$. The parameters of the simulation model are $|E_u|/|E_m|$ = 1.75, $|E_l|/|E_m|$ = 0.3, $r$ = 2.5 μm, $n_m$ = 1.5, $n_e$ = 1.0, $\lambda_u = \lambda_l = \lambda_m = \lambda_0$ = 400 nm. The calculated coordinates of start point, $I_{max}$ point and end point are (0.198 μm, 2.492 μm), (0.009 μm, 2.959 μm) and (-0.0455 μm, 4.87 μm), respectively. The bending angle is the included angle between the line connecting the $I_{max}$ point and the end point and the line connecting the $I_{max}$ point and the start point. According to the vector dot product formula in Euclidean geometry, the calculated $\delta_b$ is -20.56°.

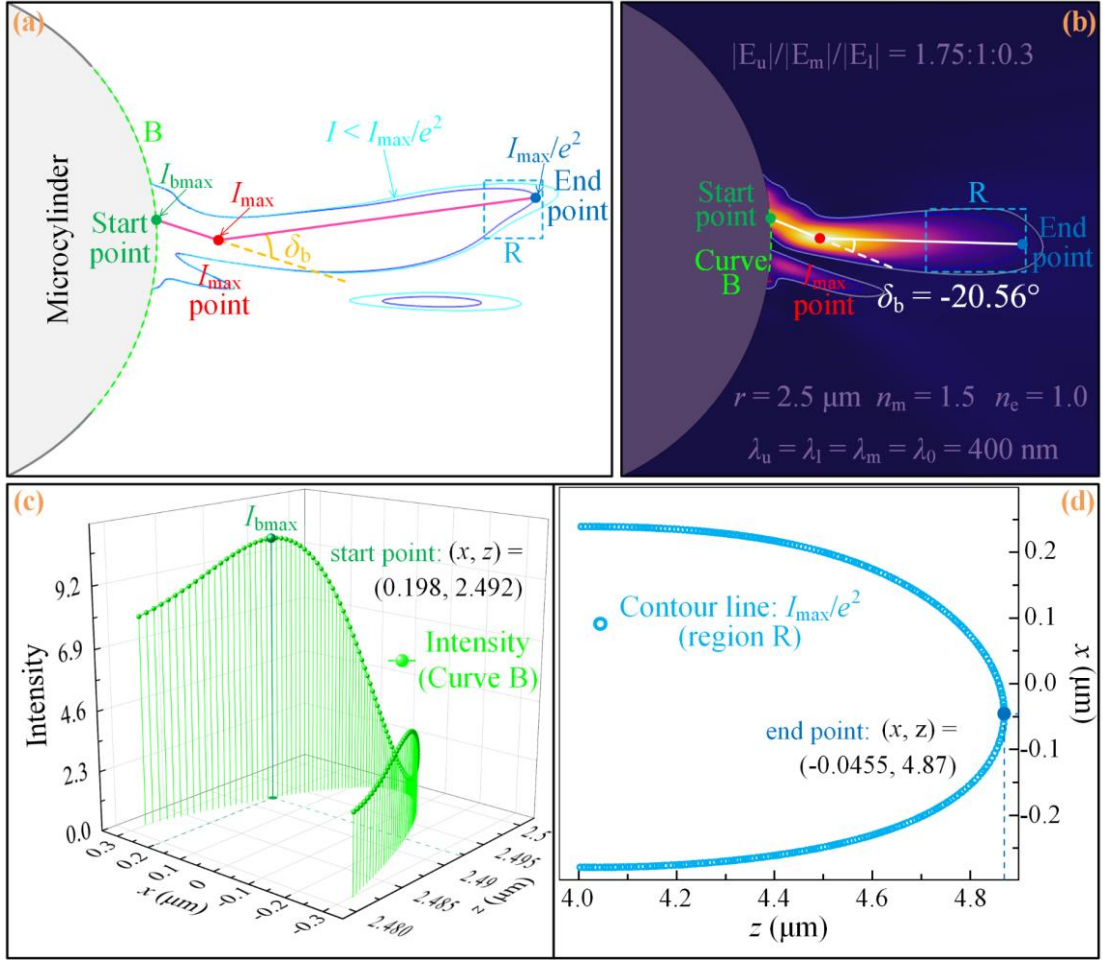

**Figure S3.** a) Schematic diagram used to define the bending angle of PH. 'B': a selected arc line on the boundary between the microcylinder and environmental medium, $I_{bmax}$: the maximum value of the light intensity distribution curve of arc line 'B', 'R': a selected rectangular area 'R' containing part of the contour curve with light intensity of $I_{max}/e^2$. The marked green, red and blue dots are respective the start point, $I_{max}$ point and end point. The angle between two pink lines is the bending angle $\delta_b$. b) Optical field distribution of an example with $|E_u|/|E_m|/|E_l|$ = 1.75:1:0.3. The radius of the microcylinder is $r$ = 2.5 μm. The refractive indices of the microcylinder and environmental medium are $n_m$ = 1.5 and $n_e$ = 1.0, respectively. The wavelength of the incident plane wave is $\lambda_u = \lambda_l = \lambda_m = \lambda_0$ = 400 nm. Light intensity distributions on c) arc line 'B' and d) $I_{max}/e^2$ contour curve within rectangular area 'R'. The highlighted 'sphere' in c) and circle in d) are the start point ($I_{bmax}$) and the end point, respectively.

## 4. An example of a PH with multiple deflection points

To illustrate multiple deflection points at a PH, the following example is established. As shown in Figure S4a, the microcylinder is divided into three parts according to the partition of RSC and C-RRC of the first LRI. The refractive index of the lower part ($n_l$) related to the C-RRC below is different from the refractive indices of the middle part ($n_m$) and the upper part ($n_u$) related to the RSC and C-RRC above. Under the irradiation of a plane wave, the PH is formed in Figure S4b. The model parameters are $n_u = n_m = 1.5$, $n_l = 1.8$, $r = 2.5$ μm and $\lambda_0 = 400$ nm. Figure S4c shows an enlarged view of the selected red rectangular area in Figure S4b. Given that it is customary to use the point of maximum light intensity to indicate where the 'inflection point' is, i.e. the maximum light intensity of the transverse optical field along $x$ direction for points at different $z$ coordinates in the direction of light propagation. The purple dots in Figure S4c are a series of points with transverse maximum light intensity obtained in this method. In the calculation, the points near the microcylinder are denser, while the points away from the microcylinder are spaced 0.1 μm apart. At least five positions, marked with yellow circles and Roman numerals from I to V, can be seen in the figure where the optical field has been deflected. Obviously, the value of the angle calculated together with the start point and the end point is not equal to the bending angle ($\delta_b = 26.89°$) given in the figure by selecting any of the five points as the 'inflection point'. Thus, we think it would be better to use the $I_{max}$ point as the unique deflection point for the definition of the bending angle.

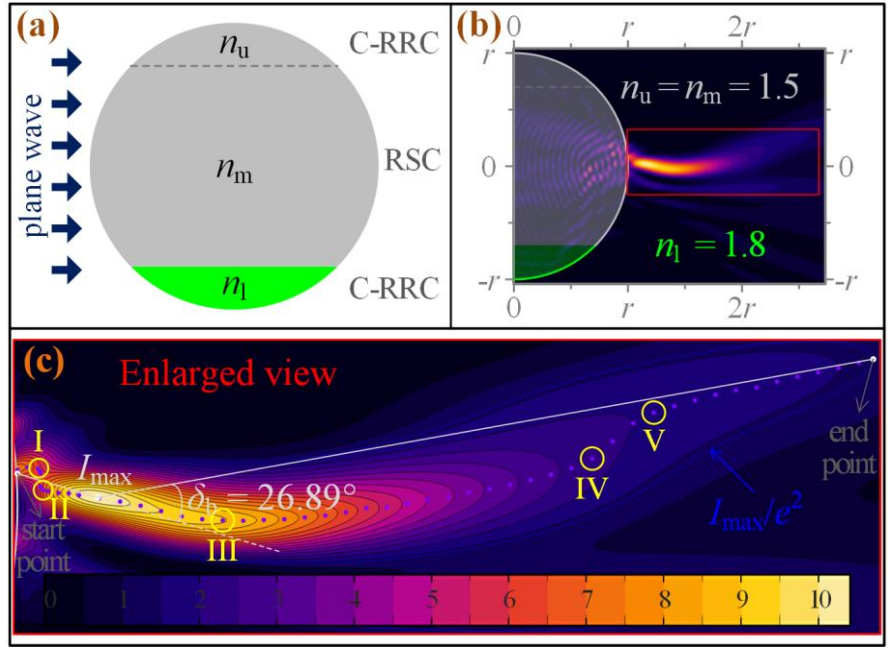

**Figure S4.** a) Schematic diagram of a plane-wave-illuminated microcylinder made of composite materials. The refractive indices of the upper, middle and lower part of the microcylinder are $n_u$, $n_m$ and $n_l$, respectively. The radius of the microcylinder is $r$. b) Optical field distribution of the PH formed by composite microcylinder at $n_u = n_m = 1.5$, $n_l = 1.8$, $r = 2.5$ μm and $\lambda_0 = 400$ nm. c) Enlarged view of the selected area in b). The yellow circles and the arguments I, II, III, IV and V marked in the figure point out five deflection points. The blue curve represents a contour line with intensity of $I_{max}/e^2$.